\documentclass[twocolumn,showpacs,amsmath,amssymb]{revtex4}

\usepackage{epsfig,psfrag}
\usepackage{dcolumn}
\usepackage{bm}
\usepackage{graphicx}
\usepackage{color}
\newcommand{\beq}{\begin{equation}}
\newcommand{\eeq}{\end{equation}}
\newcommand{\beqr}{\begin{eqnarray}}
\newcommand{\eeqr}{\end{eqnarray}}
\newcommand{\e}{{\epsilon}}
\newcommand{\s}{{\sigma}}
\newcommand{\bss}{{\bar{\sigma}}}
\newcommand{\cb}{{\bar{c}}}

\newcommand{\w}{{\omega}}

\def\tlambda{{\tilde{\lambda}}}

\def\bB{{\mathbf B}}

\def\bS{{\mathbf S}}

\def\bB{{\mathbf B}}

\def\tJ{{\tilde{J}}}

\newcommand{\sigmab}{\mbox{\boldmath $\sigma $}}

\def\half{{1\over2}}

\def\ua{\uparrow}
\def\da{\downarrow}
\def\eqa{\begin{eqnarray}}
\def\eea{\end{eqnarray}}

\def\cT{{\cal T}}

\def\a{{\alpha}}
\def\b{{\beta}}
\def\g{{\gamma}}
\def\d{{\delta}}

\def\eurlet{Europhys. Lett.}
\def\jpa{{Jour. Phys. A}}
\begin{document}

\title{Random Matrix Crossovers  and Quantum Critical Crossovers for Interacting Electrons in Quantum Dots}
\author{Ganpathy Murthy} 
\affiliation{Department of Physics and Astronomy,
University of Kentucky, Lexington KY 40506-0055} \date{\today}
\begin{abstract}
Quantum dots with large Thouless number $g$ embody a regime where both
disorder and interactions can be treated nonperturbatively using
large-$N$ techniques (with $N=g$) and quantum phase transitions can be
studied. Here we focus on dots where the noninteracting Hamiltonian is
drawn from a crossover ensemble between two symmetry classes, where
the crossover parameter introduces a new, tunable energy scale
independent of and much smaller than the Thouless energy. We show that
the quantum critical regime, dominated by collective critical
fluctuations, can be accessed at the new energy scale. The
nonperturbative physics of this regime can only be described by the
large-$N$ approach, as we illustrate with two experimentally relevant
examples.

\end{abstract}
\vskip 1cm 
\maketitle

Mesoscopic systems show many novel properties which are not present in
bulk systems\cite{mesoscopics-review}, such as the Coulomb Blockade in
the zero-bias conductance through a quanutm dot, or persistent
currents in mesoscopic rings penetrated by a magnetic flux. Disorder
renders the single-particle states chaotic\cite{qd-reviews}, with the
statistics of the energies and wavefunctions controlled by Random
Matrix Theory (RMT)\cite{rmt}, which also controls the correlations
between different states separated by less
than the Thouless energy $E_T$ (related to the ergodicization time for
a particle $\tau_{erg}$ by the Uncertainty Principle
$E_T=\hbar/\tau_{erg}$). For mean single-particle level spacing
$\delta$, $g=E_T/\delta$.

Since disorder breaks all the spatial symmetries, only time-reversal
$\cT$ and possibly Kramers degeneracy remain. There are three
classical symmetry classes\cite{rmt}, the gaussian orthogonal ensemble
or GOE ($\cT$ intact, no spin-orbit coupling), the unitary or GUE
($\cT$ broken), and the symplectic or GSE ($\cT$ intact, with
spin-orbit coupling). More recently, other classes have been
identified for disordered superconductors\cite{zirnbauer} and quantum
dots constructed from two-dimensional semiconductor heterostructures
with spin-orbit coupling\cite{aleiner-falko}.

The question of how to incorporate electron-electron interactions in
mesoscopic systems has enjoyed much attention. A recent proposal has
been (somewhat inaccurately) dubbed the ``Universal
Hamiltonian''\cite{H_U,univ-ham}. Its central idea is to start with a
general Hamiltonian
\beq H=
\sum\limits_{\a,s}\e_{\a}c^{\dagger}_{\a,s}c_{\a,s}+\half\sum\limits_{\a\b\g\d,ss'}
V^{ss'}_{\a\b\g\d}c^{\dagger}_{\a,s}c^{\dagger}_{\b,s'}c_{\g,s'}c_{\d,s}
\label{generic-ham}\eeq 
where $\a,s$ are single-particle eigenstates of the kinetic energy
(including spin or Kramers degeneracy), and $c,c^{\dagger}$ are
canonical fermion operators. For a given symmetry of the kinetic term,
only special matrix elements of the interaction have a nonzero
ensemble average\cite{H_U}. In the GOE, only $V^{ss'}_{\a\b\b\a}$,
$V^{ss'}_{\a\b\a\b}$, and $V^{s,-s}_{\a\a\b\b}$ survive ensemble
averaging. Matrix elements which do not survive ensemble-averaging are
small (of typical size $\delta/g$) as are the sample-to-sample
fluctuations of the terms which do survive\cite{H_U}. Finally, in the
large-$g$ limit all small terms are dropped\cite{H_U}, leading to the
Universal Hamiltonian
\beq
H_U=\sum\limits_{\a,s}\e_{\a}c^{\dagger}_{\a,s}c_{\a,s}+{U_0\over 2}{\hat N}^2 -J\bS^2+\lambda T^{\dagger} T
\label{univ-ham}\eeq 
where ${\hat N}$ is the total particle number, $\bS$ is the total
spin, and $T=\sum c_{\b,\da}c_{\b,\ua}$. In addition to the charging
energy, $H_U$ has an exchange energy $J$ and a superconducting
coupling $\lambda$. This last term is absent in the GUE, while the
exchange term disappears in the GSE. In the large-$g$ limit only
interaction terms which are invariant under the symmetries of the
kinetic term appear in $H_U$\cite{H_U,univ-ham}.

However, the correct way to determine if a coupling is relevant is to
carry out a renormalization group (RG) analysis\cite{rg-shankar}. If
the coupling grows under RG, it must be kept, no matter how small it
was initially, and irrespective of whether it had an ensemble average
in a particular symmetry class. Using RG, the author and Harsh Mathur
have shown that in ballistic quantum dots $H_U$ is unstable to Landau
Fermi liquid couplings for sufficiently strong
coupling\cite{qd-us1}. Using a variant of the large-$N$ approach
(which subsumes the RG\cite{qd-us2}), the author and R. Shankar have
constructed\cite{qd-us2} a controlled theory of the quantum phase
transition in the large-$g$ limit. For finite $g$ we
identified\cite{qd-us2,qd-long} a weak-coupling regime which is
controlled by $H_U$, a strong-coupling regime with a distorted Fermi
surface, and a fan-shaped {\it many-body quantum critical
regime}\cite{critical-fan} (QCR) in which the physics is dominated by
collective critical fluctuations. For weak coupling, the QCR can be
accessed at nonzero frequency or temperature $(\omega,T)>E_{QCX}$
\cite{critical-fan}, the characteristic energy scale for the crossover
in our case being\cite{qd-us2,qd-long} $E_{QCX}=r E_T$, where $r$ is a
dimensionless distance from the criticality. For weak coupling,
perturbation theory around $H_U$ is valid only for $(\omega,T)\ll
E_{QCX}$, whereas the large-$N$ approach is valid for all
$(\omega,T)\ll E_T$\cite{qd-long}. Close to the transition ($r\ll1$)
the large-$N$ approach is superior, but is superfluous deep in the
weak-coupling regime ($r\simeq 1$).

The focus of this paper is a class of problems where the QCR can be
accessed at a tunable energy independent of, and much smaller than
$E_T$, even at weak-coupling, and where the large-$N$ approach is
essential to capturing the physics. These problems occur in systems
where the single-particle Hamiltonian is undergoing a crossover from
one symmetry class to another, say from the GOE to the GUE\cite{rmt},
acheived by turning on the orbital effects of a magnetic field. For
Thouless number $g\gg1$ of the original GOE, the crossover Hamiltonian
is a $g\times g$ matrix
\beq
H_{X}(\a)= H_{GOE}+{\a\over \sqrt{g}} H_{GUE}
\label{hcross}\eeq
where $\a$ is the crossover parameter. Properties of single
eigenvector\cite{cross-single}, and more recently multiple
eigenvector\cite{adam-x} correlations, have been computed in the
crossover. For $2\a^2=g_X\gg1$, the following ensemble-averaged
correlations hold for the eigenstates $\psi_{\mu}(i)$, where
$\mu\ne\nu$ label the states and $i,j,k,l$ the original orthogonal
labels:
\beqr
\langle\psi^*_{\mu}(i)\psi_{\nu}(j)\rangle =&{1\over g} \delta_{\mu\nu}\delta_{ij}\nonumber \\
\langle\psi^*_{\mu}(i)\psi^*_{\nu}(j)\psi_{\mu}(k)\psi_{\nu}(l)\rangle=&{\delta_{ik}\delta_{jl}\over g^2}+{\delta_{ij}\delta_{kl}\over g^2} {{E_{X}}\delta/\pi\over {E_{X}}^2+ (\e_{\mu}-\e_{\nu})^2}
\label{cross-correlations}\eeqr
The last term on the second line shows the extra correlations induced
in the crossover\cite{adam-x}. The energy scale ${E_{X}}=g_X\delta/\pi$
represents a window within which GUE correlations have spread. When
${E_{X}}\approx E_T$, the crossover is complete. Similar expressions hold
for the GOE to GSE crossover as well\cite{adam-x}.

It was pointed out\cite{adam-x} that interaction matrix elements have
enhanced sample-to-sample fluctuations in crossover ensembles,
controlled by $1/g_X$ and not $1/g$. Thus a naive application of the
symmetry principle leading to the Universal Hamiltonian is
problematic\cite{adam-x,gorokhov}. 

\begin{figure}[h]
\includegraphics*[width=2.4in,angle=0]{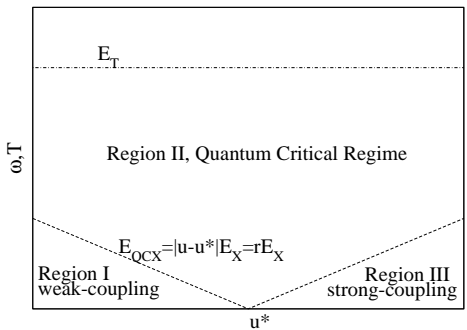}
\caption{Different regimes and crossover energy scales}
\label{fig1}
\end{figure}

We are now ready to state our central results. Consider phase
transitions driven by interactions which commute with the symmetries
of the initial ensemble but not with those of the crossover
ensemble. Then, (i) The energy scale $E_{QCX}$ characterizing the
many-body weak-coupling to quantum-critical crossover is proportional
to ${E_{X}}$. (ii) The large-$N$ approach gives a nonperturbative
description of the physics valid for energies $(\omega,T)\ll E_T$
(regions I and II in Fig.\ref{fig1}), while perturbative treatments
based on $H_U$ are valid\cite{qd-long} only for $(\omega,T)\ll
E_{QCX}\approx {E_{X}}$ (region I). (iii) Tuning the single-particle
crossover offers a powerful way to control and access the
quantum-critical crossover. These results apply equally to diffusive
and ballistic dots. Finally, RMT is much better controlled in the
crossover regime ${E_{X}}\ll E_T$\cite{nowindow} than when applied to
the entire Thouless shell\cite{qd-us2,qd-long}.

The tunability of the QCR is potentially very important for quantum
dots fabricated from two-dimensional semiconductor heterostructures,
since current samples appear to be on the weak-coupling side of the
Stoner and other transitions (see references in
refs.\cite{univ-ham,qd-long}). An illustration relevant for these
samples is provided by the crossover from the GOE to the new
``intermediate'' symplectic ensemble\cite{aleiner-falko} with small
spin-orbit coupling in two-dimensional heterostructures.  Here the
physical crossover parameter $\a$ is a function of $L/L_{SO}$, where
$L_{SO}$ is the spin-orbit scattering length. Keeping the lowest
leading order in this ratio ($\a=CL^2/L_{SO}^2$, where $C$ is a factor
of order unity) leads\cite{aleiner-falko} to a Hamiltonian which
conserves $S_z$, but not the total spin. The Universal Hamiltonian for
this case has recently been worked out by Alhassid and
Rupp\cite{alhassid-rupp} (AR). For the sake of simplicity, let us
ignore all interactions except exchange. The Hamiltonian can be
written as
\beq
H=\sum \e_{\mu s}c^{\dagger}_{\mu s} c_{\mu s} -J_z S_z^2 -J(S_x^2+S_y^2)
\eeq
The last term would be absent in the Universal
Hamiltonian\cite{alhassid-rupp}, since it does not commute with the
kinetic term, but physically it should be present because the
high-energy processes which led to it are still
present. AR\cite{alhassid-rupp} show that the spin operators can be
written as
\beq
S_+=\sum\limits_{\mu\nu} M^*_{\nu\mu} c^{\dagger}_{\mu,\ua}c_{\nu,\da}
\eeq
with $S_-=(S_+)^{\dagger}$. The quantities $M_{\mu\nu}$ satisfy
\beq
\langle|M_{\mu\nu}|^2\rangle={{E_{X}}\delta/\pi\over {E_{X}}^2+(\e_{\mu,\ua}-\e_{\nu,\da})^2}
\eeq
where ${E_{X}}=C^2\a^2\pi E_T$ (similar to Eq.
(\ref{cross-correlations})). It is straightforward to carry out the
Hubbard-Stratanovich transformation, integrate out fermions, and
obtain an effective action for the collective variables, here $\s_x$,
$\s_y$ coupling to $S_x$, $S_y$. Deep in the crossover $E_X\gg\delta$
the effective action is self-averaging\cite{qd-us2,qd-long}.  Calling $J=\tJ
\delta$, and taking $g\to\infty$ and setting $J_z=0$ for simplicity, to quadratic order  we have the Euclidean action
\beq S_{eff}=\delta\int {d\omega\over 2\pi} |{\vec\s}(\omega)|^2
({1\over\tJ}-{1\over1+|\omega|/{E_{X}}})
\label{seff-j}\eeq 
The critical coupling is $\tJ^*=1$, which
is also the bulk critical value, exactly as in ref.\cite{nowindow} (in
fact, the quadratic effective action for the model of
ref.\cite{nowindow}, which the authors did not compute, would be
identical to Eq. (\ref{seff-j})). The retarded $S_+S_-$ correlation
function in weak-coupling ($\tJ<1$) at $T=0$ is read off by making the
replacement $i\omega\to\omega+i0^+$ in Eq. (\ref{seff-j}). The
imaginary part of this correlator has the scaling form valid for
$\omega\ll E_T$
\beq
-{1\over\pi} Im(\langle0|S_+(\omega)S_-(-\omega)|0\rangle)= {\tJ^2\over\pi r}{x\over1+x^2}
\label{ims+s-}\eeq
where $r=1-\tJ$ and $x=\omega/{E_{X}}r$ is the scaling variable. For
$\omega\gg {E_{X}}r\equiv E_{QCX}$ the system crosses over to the
QCR. Eq. (\ref{ims+s-}) illustrates all our central results, the fact
that $E_{QCX}\propto E_X$, the fact that the large-$N$
approach\cite{qd-us2,qd-long} captures the nonperturbative physics of
collective critical  fluctuations, and the tunability of the crossover
to the QCR.

To illustrate the generality of these results, consider the GOE to GUE
crossover in three-dimensional ballistic/chaotic superconducting
nanoparticles of linear size $L$. The single-particle spacing is
$\delta\approx
\hbar^2/mk_FL^3$ and the Thouless energy is $E_T\approx
\hbar^2k_F/mL$, leading to a Thouless number $g\approx (k_FL)^2$. A
commonly-used model to understand superconductivity in such
particles\cite{ralph-vondelft} in the absence of an orbital magnetic
field is the reduced BCS Hamiltonian
\beq
H_{BCS}=\sum\e_i c^{\dagger}_{i,s}c_{i,s}-\delta\tlambda T^{\dagger} T
\label{hbcs}\eeq
where ${i,s}$ are orthogonal state and spin labels, $\tlambda>0$ is
the attractive dimensionless $BCS$ coupling valid in an energy shell
of width $2\hbar\omega_D$ around the Fermi energy, and $T=\sum
c_{i,\da}c_{i,\ua}$.  Much recent work has concentrated on {\it
ultrasmall} grains\cite{ralph-vondelft} of size a few nanometers,
where $\delta$ is comparable to the bulk BCS gap
$\Delta=2\hbar\omega_D\exp{-1/\tlambda}$. The orbital effects of an
external magnetic field can be captured by the crossover Hamiltonian
of Eq. (\ref{hcross}), with $\a/\sqrt{g}=C(\phi/\phi_0)$, where
$\phi=BL^2$ is the flux through the sample, $\phi_0=h/e$ is the flux
quantum, and $C$ is a constant of order unity. This leads to
${E_{X}}={2C^2\over \pi} \bigg({\phi\over\phi_0}\bigg)^2 E_T\propto B^2L^3$.
For grains of $Pb$ or $Al$ smaller than 10$nm$ and fields in the
sub-Tesla range this crossover scale is smaller than $\delta$, and
thus orbital magnetic effects are of no importance for their
physics. However, for grains in the size range 10-30$nm$ in moderate
fields (see, for example, ref. \cite{reich}), the orbital effects will
be relevant, while the Zeeman coupling can be ignored. We will assume
that the critical temperature of the superconductor is smaller than
the Thouless energy of the particle, true of $Pb$ or $Al$ particles in
this size regime. Also, $L$ should be smaller than the London
penetration depth. In the presence of an orbital $\bB$ field, the
kinetic energy of Eq. (\ref{hbcs}) will be replaced by a random matrix
drawn from the ensemble of Eq. (\ref{hcross}).  Since the high-energy
processes which led to an attractive BCS interaction are not modified
by the small orbital field, the BCS interaction must be kept, though
it no longer commutes with the symmetries of the kinetic energy. This
leads to
\beqr
&H_{BCSX}=\sum\e_{\mu} c^{\dagger}_{\mu,s}c_{\mu,s}-\delta\tlambda T^{\dagger} T\\
T=&\sum\limits_{\mu\nu}M_{\mu\nu} c_{\nu,\da}c_{\mu,\ua}:\ \ \ 
M_{\mu\nu}=\sum\limits_{i} \psi_{\mu}(i)\psi_{\nu}(i)
\label{hbcsx}\eeqr
To study the magnetization of the particle in the crossover we start
with the partition function $Z=Tr(exp{-\b H})$ where $\b=1/T$ is the
inverse temperature. We convert the partition function into an
imaginary time path integral and use the Hubbard-Stratanovich identity to
decompose the interaction, leading to the imaginary time Lagrangian
\beq 
{\cal L}={|\s|^2\over\delta\tlambda}-\sum\limits_{\mu,s}\cb_{\mu,s}(\partial_{\tau}-\e_{\mu})c_{\mu,s}+\s{\bar T}+\bss T
\eeq 
where $\s,\bss$ are the bosonic Hubbard-Stratanovich fields
representing the BCS order parameter and $\cb,c$ are Grassman fields
representing fermions.  The fermions are integrated out, and the
resulting action for $\s,\bss$ is expanded to second order to
obtain
\beqr 
S_{eff}\approx {\delta\over\b}\sum\limits_{n}& |\s(i\omega_n)|^2
({1\over\tlambda}-f_n(\b,{E_{X}},\omega_D)) \\
f_n(\b,{E_{X}},\omega_D)=&\delta\sum\limits_{\mu\nu} |M_{\mu\nu}|^2
{1-N_F(\e_{\mu})-N_F(\e_{\nu})\over \e_{\mu}+\e_{\nu}-i\omega_n} 
\eeqr
where $\omega_n=2\pi n/\b$, and the sums are restricted to
$|\e_{\mu}|,|\e_{\nu}|<\hbar\omega_D$. We see that the correlations
between different states $\mu,\nu$ play an important role. Deep in the
crossover (for ${E_{X}}\gg\delta$) we can replace
$|M_{\mu\nu}|^2$ by its ensemble average (dominated by the last term
of Eq. (\ref{cross-correlations})), just as in our previous
work\cite{qd-us2,qd-long}. We will also henceforth replace the
summations over energy eigenstates by energy integrations with the
appropriate cutoffs. The sums can be approximately carried out to give
\beq 
f_n\approx \half\log\bigg({4(\hbar\omega_D)^2+\omega_n^2\over
 C'/\b^2+({E_{X}}+|\omega_n|)^2}\bigg) 
\eeq 
where $C'\approx3.08$ is chosen to obtain the correct transition
temperature in the absence of a magnetic field.  The coefficient of
the quadratic term for the $i\omega_n=0$ term as $T\to 0,\
\b\to\infty$ is ${1\over\tlambda}-\log{{2\hbar\omega_D\over {E_{X}}}}$,
leading to a continuous phase transition at
$\tlambda^*=1/\log(2\hbar\omega_D/E_X)$.

On the normal side ($\tlambda<\tlambda^*$) $\s,\bss$ fluctuate, and
the quadratic action is sufficient to explore their fluctuations. One
can now integrate out $\s,\bss$ to obtain the contribution of their
thermal fluctuations to the free energy. One then obtains for the
magnetization
\beq
M=-{\partial F\over\partial B}=M_{nonint}+\tlambda L^2{d{E_{X}}\over
d\phi}\sum\limits_{n} {{\partial f_n\over\partial
{E_{X}}}\over1-\tlambda f_n}
\eeq
where $M_{nonint}$ is the contribution to the magnetization from
noninteracting electrons (intimately connected to the noninteracting
persistent current\cite{mesoscopics-review}).

To see the physics of the crossover into the QCR as $T$
increases\cite{critical-fan} in the most transparent way, we take the
weak-coupling limit $\tlambda\to0$, $\hbar\omega_D\to\infty$ such that
$T_c$ remains invariant. Defining
$r={\tlambda}^{-1}-(\tlambda^*)^{-1}$, we obtain for the fluctuation
magnetization in the scaling region
\beq
-M_0\sum\limits_{n=-\infty}^{\infty} {x(1+2\pi |n|x)/(C'x^2+(1+2\pi
 |n|x)^2)\over r+\half\log(C'x^2+(1+2\pi |n| x)^2)}
\label{mfluc}\eeq
where $x=T/{E_{X}}$ is the scaling variable, and
$M_0=L^2\sqrt{(8C^2{E_{X}}E_T/\pi\phi_0^2)}$ sets the $T$-independent
scale for the magnetization. At quantum criticality $r=0$. For $r>0$
and $T\gg {E_{X}}$ the second log in the denominator dominates, and
the system crosses over into the QCR. In this case, unlike the
previous one, the critical coupling $\tlambda^*$ depends on $E_{X}$,
making the critical point tunable as well.  Note that this result is
nonperturbative in ${E_{X}}$, $T$, and $\tlambda$, and offers a
perspective complementary to previous results which are perturbative
in one or more of these
parameters\cite{ambegaokar,schmid,schechter}. The regime of validity
of Eq. (\ref{mfluc}) is $E_T\gg (T,{E_{X}})\gg\delta$,  and $r>0$.

In summary, we have identified a class of mesoscopic problems in which
access to the quantum critical regime (where the physics is dominated
by collective critical fluctuations) can be obtained at energies
$E_{QCX}$ much smaller than $E_T$ by tuning the crossover scale $E_X$
between ensembles with different single-particle symmetry.
Perturbative approaches in these systems are limited to $(\omega,T)\ll
E_{QCX}\approx {E_{X}}$, whereas the large-$N$
approach\cite{qd-us2,qd-long} is successful in capturing their
nonperturbative physics in the entire range $(\omega,T)\ll
E_T$. Symmetries crossovers in weak-coupling large-$g$ dots offer
theoretical control over and experimental access to strongly
correlated physics. It would be interesting (and directly relevant to
superconducting particles) to see the crossover from the
strong-coupling regime to the quantum critical regime, as well as to
determine the signatures of the weak-coupling to quantum critical
crossover in the transport properties of semiconductor quantum dots.

It is a pleasure to thank R. Shankar and Y. Alhassid for illuminating
conversations, and the NSF for partial support under DMR-0311761.


\begin{thebibliography}{99}
\bibitem{mesoscopics-review} For reviews see, Y. Imry, {\it
Introduction to Mesoscopic Physics}, Oxford University Press,
1997; K. B. Efetov, {\it Supersymmetry in Disorder and Chaos},
Cambridge University Press, 1997.

\bibitem{qd-reviews} For recent reviews, see, T. Guhr, A. M\"uller-Groeling,
and H. A. Weidenm\"uller, Phys. Rep. {\bf 299}, 189 (1998); Y.
Alhassid, \rmp\ {\bf 72}, 895 (2000); A. D. Mirlin, Phys. Rep.
{\bf 326}, 259 (2000)

\bibitem{rmt} M. L. Mehta, {\it Random Matrices}, Academic Press, San
Diego, 1991.

\bibitem{zirnbauer} M. R. Zirnbauer, J. Math. Phys. {\bf 37}, 4986 (1996).

\bibitem{aleiner-falko} I. L. Aleiner and V. I. Fal'ko, \prl {\bf 87}, 256801 (2001); {\bf 89}, 079902 (E) (2002). 

\bibitem{H_U} A.  V.  Andreev and A.
Kamenev, \prl {\bf 81}, 3199 (1998); P.  W.  Brouwer, Y.  Oreg,
and B. I.  Halperin, \prb {\bf 60}, R13977 (1999); H.  U.
Baranger, D. Ullmo, and L.  I.  Glazman, \prb {\bf 61}, R2425
(2000); I.  L. Kurland, I.  L.  Aleiner, and B.  L.  Al'tshuler,
\prb\ {\bf 62}, 14886 (2000).

\bibitem{univ-ham}I. L. Aleiner, P. W. Brouwer, and L. I. Glazman,
Phys. Rep. {\bf 358}, 309 (2002), and references therein; Y. Oreg,
P. W. Brouwer, X. Waintal, and B. I. Halperin, cond-mat/0109541,
and references therein.

\bibitem{rg-shankar} R. Shankar, {\it Physica}\ {\bf A177},
530 (1991); R.Shankar, { Rev. Mod. Phys.} {\bf 66}, 129 (1994).


\bibitem{qd-us1} G. Murthy and H. Mathur, \prl\ {\bf 89}, 126804 (2002).

\bibitem{qd-us2} G. Murthy and R. Shankar, \prl {\bf 90}, 066801 (2003).

\bibitem{qd-long} G. Murthy, R. Shankar, D. Herman, and H. Mathur, \prb 
{\bf 69}, 075321 (2004).

\bibitem{critical-fan} S. Chakravarty, B. I. Halperin, and
D. R. Nelson, \prl\ {\bf 60}, 1057 (1988); \prb\ {\bf 39}, 2344
(1989); For a detailed treatment of the generality of the
phenomenon, see, S. Sachdev, {\it Quantum Phase Transitions},
Cambridge University Press, Cambridge 1999.

\bibitem{cross-single} H. -J. Sommers and S. Iida, Phys. Rev. E {\bf 49}, 
2513 (1994); V. I. Fal'ko and K. B. Efetov, \prb {\bf 50}, 11267
(1994); \prl {\bf 77}, 912 (1996); S. A. van Langen, P. W. Brouwer,
and C. W. J. Beenakker, Phys. Rev. E {\bf 55}, 1 (1997).

\bibitem{adam-x} S. Adam, P. W. Brouwer, J. B. Sethna, 
and X. Waintal, \prb {\bf 66}, 165310 (2002).

\bibitem{gorokhov} D. A. Gorokhov and P. W. Brouwer, \prl {\bf 91}, 
186602 (2003); \prb {\bf 69}, 155417 (2004).

\bibitem{nowindow} S. Adam, P. W. Brouwer, and P. Sharma, \prb {\bf 68}, 
241311 (2003).

\bibitem{alhassid-rupp} Y. Alhassid and T. Rupp, cond-mat/0312691 (2003). 

\bibitem{ralph-vondelft} J. von Delft and D. C. Ralph, Phys. Rep. {\bf 345}, 
61 (2001); J. von Delft, Ann. Phys. (Leipzig) {\bf 10}, 1 (2001).

\bibitem{reich} S. Reich, G. Leitus, R. Popovitz-Biro, and M. Schechter, 
\prl {\bf 91}, 147001 (2003). 

\bibitem{ambegaokar} V. Ambegaokar and U. Eckern, \prl {\bf 65}, 381 (1990); 
Europhys. Lett. {\bf 13}, 733 (1990).

\bibitem{schmid} A. Schmid, \prl {\bf 66}, 80 (1991).

\bibitem{schechter} M. Schechter, Y. Oreg, Y. Imry, and Y. Levinson, \prl {\bf 90}, 026805 (2003). 



\end{thebibliography}
\end{document}